\documentclass[aps,prb,twocolumn,showpacs,superscriptaddress]{revtex4-1}
\usepackage{amsmath}
\usepackage{graphicx}
\usepackage{subfigure}
\usepackage{ulem,color}

\begin{document}
\title{Boosting Domain Wall Propagation by Notches }

\author{H. Y. Yuan and X. R. Wang}
\email{Corresponding author: phxwan@ust.hk}
\affiliation{Physics Department, The Hong Kong University of Science
and Technology, Clear Water Bay, Kowloon, Hong Kong}
\affiliation{HKUST Shenzhen Research Institute, Shenzhen 518057, China}

\begin{abstract}
We report a counter-intuitive finding that notches in an otherwise homogeneous
magnetic nanowire can boost current-induced domain wall (DW) propagation.
DW motion in notch-modulated wires can be classified into three phases:
1) A DW is pinned around a notch when the current density is below the
depinning current density. 2) DW propagation velocity is boosted
by notches above the depinning current density and when non-adiabatic
spin-transfer torque strength $\beta$ is smaller than the Gilbert
damping constant $\alpha$. The boost can be manyfold.
3) DW propagation velocity is hindered when $\beta > \alpha$.
The results are explained by using the Thiele equation.
\end{abstract}

% 75.60.Ch domains and domain wall structure
% 75.78.Cd micromagnetic simulation
% 85.70.Ay magnetic device characterization, design, and modeling
% 85.70.Kh magnetic thin film devices
\maketitle

\section{Introduction}
Magnetic domain wall (DW) motion along a nanowire underpins
many proposals of spintronic devices \cite{Parkin2008,Cowburn}.
High DW propagation velocity is obviously important because it
determines the device speed. In current-driven DW propagation,
many efforts have been devoted to high DW velocity and low
current density in order to optimize device performance.
%Recently, electric fields and strains have been demonstrated
%to be effective DW control knobs.
%\cite{Schellekens,Chiba,Lei,Bauer,Tetienne,Ranieri}.
The issue of whether notches can enhance current-induced
DW propagation is investigated here.

Traditionally, notches are used to locate DW positions
\cite{Parkin2008,Cowburn,Klaui2005,Hayashi2006}.
Common wisdom expects notches to strengthen DW pinning and to
hinder DW motion. Indeed, in the field-driven DW propagation,
intentionally created roughness slows down DW propagation
although they can increase the Walker breakdown field \cite{Milta}.
Unlike the energy-dissipation mechanism of field-induced
DW motion \cite{Wang2009}, spin-transfer torque (STT)
\cite{Berger1984, Slonczewski1996,Zhang2004,Thiaville2005}
is the driven force behind the current-driven DW motion.
The torque consists of an adiabatic STT and a much smaller
non-adiabatic STT \cite{Zhang2004,Thiaville2005}.
In the absence of the non-adiabatic STT, there exists an intrinsic
pinning even in a homogeneous wire, below which a sustainable
DW motion is not possible \cite{Li2004, Tatara2004}.
Interestingly, there are indications \cite{Yuan2014} that the depinning
current density of a DW trapped in a notch is smaller than the intrinsic
threshold current density in the absence of the non-adiabatic STT. 
Although there is no intrinsic pinning \cite{Thiaville2005,Parkin2008}
in the presence of a non-adiabatic STT,
It is interesting to ask whether notches can boost DW propagation
in the presence of both adiabatic STT and non-adiabatic STT.

In this paper, we numerically study how DW propagates along
notch-modulated nanowires. Three phases are identified: pinning
phase when current density is below depinning current density $u_d$;
boosting phase and hindering phase when the current density is above
$u_d$ and the non-adiabatic STT strength $\beta$ is smaller or
larger than the Gilbert damping constant $\alpha$, respectively.
The average DW velocity in boosting and hindering phases is
respectively higher and lower than that in the wire without notches.
It is found that DW depinning is facilitated by antivortex nucleation.
In the case of $\beta<\alpha$, the antivortex generation is responsible
for velocity boost because vortices move faster than transverse walls.
In the other case of $\beta>\alpha$, the longitudinal velocity of a
vortex/antivortex is slower than that of a transverse wall in a
homogeneous wall and notches hinder DW propagation.

\section{Model and Method}
\begin{figure}
\centering
\includegraphics[width=0.48\textwidth]{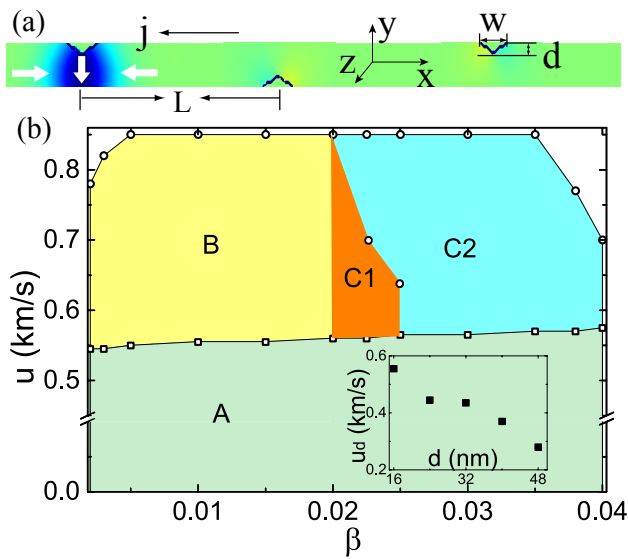}\\
\caption{(color online) (a) A notch-modulated nanowire.
$L$ is the separation between two adjacent notches.
The color codes the $y-$component of $\mathbf{m}$ with red for $m_y=1$,
blue for $m_y=-1$ and green for $m_y=0$.
The white arrows denote magnetization direction.
(b) The phase diagram in $\beta-u$ plane.
%(top-left boundary of B and top-right boundary of C): A is the
A is the pinning phase; B is the boosting phase;
and C is the hindering phase.
Vortices are (are not) generated near notches by a propagating DW in
C1 (C2). Inset: The notch depth dependence of depinning current
$u_d$ when notch width is fixed at $w=48$ nm.
}
\label{fig1}
\end{figure}

We consider sufficient long wires (with at least 8 notches) of various 
thickness and width. It is well known that \cite{McMichael} narrow 
wires favor only transverse walls while wide wires prefer vortex walls. 
Transverse walls are the main subjects of this study.
A series of identical triangular notches of depth $d$ and width $w$
are placed evenly and alternately on the two sides of
the nanowires as shown in Fig. \ref{fig1}a with a typical clockwise
transverse wall pinned at the center of the first notch.
The $x-$, $y-$ and $z-$axis are along length, width, and thickness
directions, respectively. The magnetization dynamics of the wire is
governed by the Landau-Lifshitz-Gilbert (LLG) equation
\begin{equation*}
\frac{\partial\mathbf{m}}{\partial t} = - \gamma
\mathbf{m \times H}_{\mathrm{eff}} +
\alpha \mathbf{m} \times \frac{\partial\mathbf{m}}{\partial t} -
(\mathbf{u} \cdot \mathbf{\nabla} )\mathbf{m}+\beta \mathbf{m}
\times (\mathbf{u} \cdot \mathbf{\nabla} )\mathbf{m},
\end{equation*}
where $\mathbf{m}$, $\gamma$, $\mathbf{H}_{\mathrm{eff}}$, and
$\alpha$ are respectively the unit vector of local magnetization,
the gyromagnetic ratio, the effective field including exchange and
anisotropy fields, and the Gilbert damping constant.
The third and fourth terms on the right hand side are the adiabatic
STT and non-adiabatic STT \cite{Thiaville2005}. The vector $\mathbf{u}$
is along the electron flow direction and its magnitude is $u=jP\mu_B
/(eM_s)$, where $j$, $P$, $\mu_B$, $e$, and $M_s$ are current density,
current polarization, the Bohr magneton, the electron charge and the
saturation magnetization, respectively. For permalloy of $M_s =8 \times
10^5$ A/m, $u=100$ m/s corresponds to $j=1.4\times 10^{12}\ \mathrm{A/m^2}$.
In this study, $u$ is limited to be smaller than both 850 m/s
(corresponding to $j\simeq 1.2\times 10^{13}\ \mathrm{A/m^2}$!) and
the Walker breakdown current density because current density above
the values generates intensive spin waves around DWs and notches,
which makes DW motion too complicated to be even described.
Dimensionless quantity $\beta$ measures the strength of non-adiabatic STT
and whether $\beta$ is larger or smaller than $\alpha$ is still in debate
\cite{Thiaville2005,Beach2006,Parkin2010}. The LLG equation is numerically
solved by both {\footnotesize OOMMF} \cite{oommf} and {\footnotesize MUMAX}
\cite{mumax} packages \cite{note1}.
The electric current density is modulated according to wire cross section
area while the possible change of current direction around notch is neglected.
The material parameters are chosen to mimic permalloy with exchange stiffness
$A=1.3 \times 10^{-11}$ J/m, $\alpha = 0.02$ and $\beta$ varying from
$0.002$ to $0.04$. The mesh size is $4 \times 4 \times 4\ \mathrm{nm^3}$.

\section{Results}
\subsection{Transverse walls in wide wires: boosting and hindering}
This is the focus of this work. 
Our simulations on wires of 4 nm thick and width ranging from 32 nm to 
128 nm and notches of $d=16$ nm and $w$ varying from 16 nm to 128 nm 
show similar behaviors. Domain walls in these wires are transverse. 
Results presented below are on a wire of 64 nm wide and
notches of $w=48$ nm. Three phases can be identified. A DW is pinned
at a notch when $u$ is below a depinning current density $u_d$.
This pinning phase is denoted as A (green region) in Fig. 1b.
Surprisingly, $u_d$ increases slightly with $\beta$, indicating that
the $\beta$-term actually hinders DW depinning out of a notch although
it is responsible for the absence of the intrinsic pinning in a uniform
wire (see discussion below for possible cause).
When $u$ is above $u_d$, a DW starts to propagate and it can
either be faster or slower than the DW velocity in the corresponding
uniform wire, depending on relative values of $\beta$ and $\alpha$.

When $\beta < \alpha$, DW velocity is boosted through antivortex generation
at notches.
This phase is denoted as phase B. When $\beta > \alpha$, the
boosting of DW propagation is suppressed no matter vortices are generated
(phase C1) or not (phase C2). The upper bound of the phase plane is
determined by the Walker breakdown current density and $u=850$ m/s.
If the current density is larger than the upper bound, spin waves emission
from DW \cite{hubin} and notches are so strong that new DWs may be created.
Also, the Walker breakdown is smaller than the depinning value $u_d$ for
$\beta>0.04$. Thus the phase plane in Fig. 1b is bounded by $\beta=0.04$.
Although the general phase diagram does not change, the phase boundaries
depend on the wire and notch specificities.
The inset is notch depth dependence of the depinning current
when $w=48$ nm and $\beta=0.01$ \cite{note2}.

\begin{figure}
\centering
\includegraphics[width=0.48\textwidth]{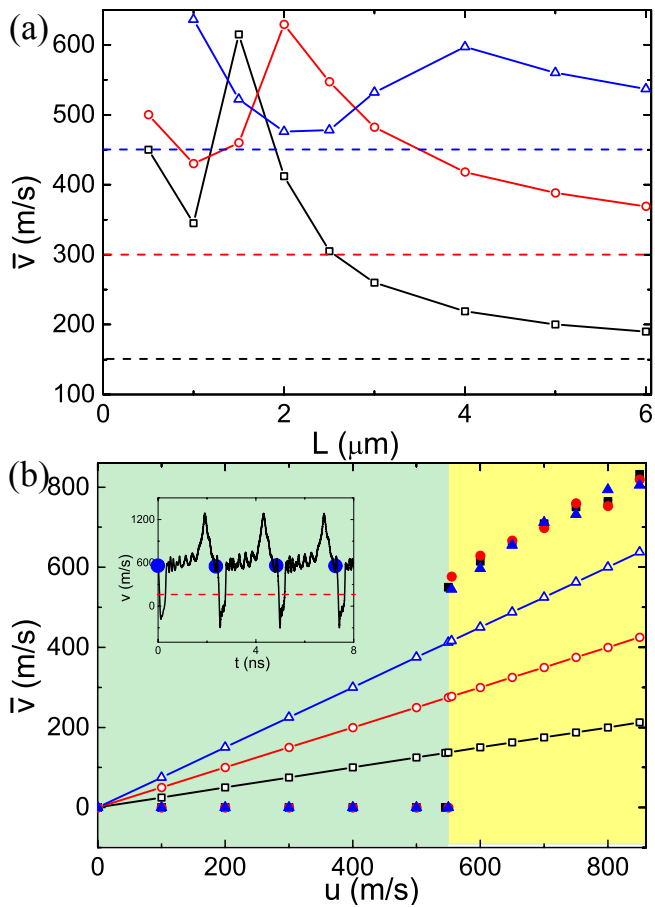}\\
\caption{(color online) (a) $L-$dependence of average DW velocity $\bar
v$ for $u=600$ m/s, $\alpha =0.02$, and $\beta=0.005$ (squares), 0.01
(circles), 0.015 (up-triangles). The dash lines are $\beta u/\alpha$.
(b) $u-$dependence of $\bar v$ for $\beta=0.005$ (squares), 0.01
(circles), 0.015 (up-triangles). Open symbols are DW velocity in the
corresponding homogeneous wires. Straight lines are $\beta u/\alpha$.
$\bar v$ is above $\beta u/\alpha$ when $u>u_d$.
Inset: instantaneous DW speed for $u=600$ m/s, $\beta=0.005$, and
$L=1.5\  \mu$m.
The blue dots indicate the moments when the DW is at notches.}
\label{fig2}
\end{figure}

\noindent\underline{Boosting phase:}
The boost of DW propagation for $\beta < \alpha$ can be clearly
seen in Fig. 2. Figure \ref{fig2}a is the average DW velocity $\bar
v$ as a function of notch separation $L$ for $u=600$ m/s $>u_d$.
$\bar v$ is maximal around an optimal notch separation $L_p$, which is close 
to the longitudinal distance that an antivortex travels in its lifetime.
$L_p$ increases with $\beta$ and it is respectively about 1.5 $\mu$m,
2 $\mu$m, and 4 $\mu$m for $\beta=0.005$ (squares), 0.01 (circles) and 
0.015 (up-triangles). This result suggests that the antivortex generation 
and vortex dynamics are responsible for the DW propagation boost. 
Filled symbols in Fig. \ref{fig2}b are $\bar v$ for various current 
density when $L_p$ is used. For a comparison, DW velocities in the 
corresponding homogeneous wires are also plotted as open symbols 
which agree perfectly with $\bar v=\beta u/\alpha$ discussed below.
Take $\beta=0.005$ as an example, $\bar v$ is zero below $u_d=550$ m/s
and jumps to an average velocity $\bar v\simeq 550$ m/s at $u_d$,
which is about four times of the DW velocity in the homogeneous wire.
As the current density further increases, the average velocity also
increases and is approximately equal to $u$.
The inset of Fig. \ref{fig2}b shows the instantaneous DW velocity for
$\beta=0.005$ and $u=600$ m/s. Blue dots denote the moments at which the
DW is at notches. Right after the current is turned on at $t=0$ ns, the
instantaneous DW velocity is very low until an antivortex of winding number
$q=-1$ \cite{Chern,Yuan_jmmm2014} is generated near the notch edge at
0.5 ns (see discussion and Fig. 9 below). The motion of the antivortex core
drags the whole DW to propagate forward at a velocity around 600 m/s.
The antivortex core annihilates itself at the bottom edge of the wire
after traveling about 1.5 $\mu$m and the initial transverse wall
reverses its chirality at the same time \cite{Yuan_jmmm2014}.
Surprisingly, the reversal of DW chirality leads to a significant
increases of DW velocity as shown by the peaks of the
instantaneous velocity at about 2.0ns in the inset. 
Another antivortex of winding number $q=-1$ 
is generated at the second notch and DW propagation speeds up again.
Once the antivortex core forms, it pulls the DW out of notch.
This process then repeats itself and the DW propagates at an
average longitudinal velocity of about 600 m/s.
A supplemental movie corresponding to the inset is attached \cite{Movie}.

\begin{figure}
\centering
\includegraphics[width=0.48\textwidth]{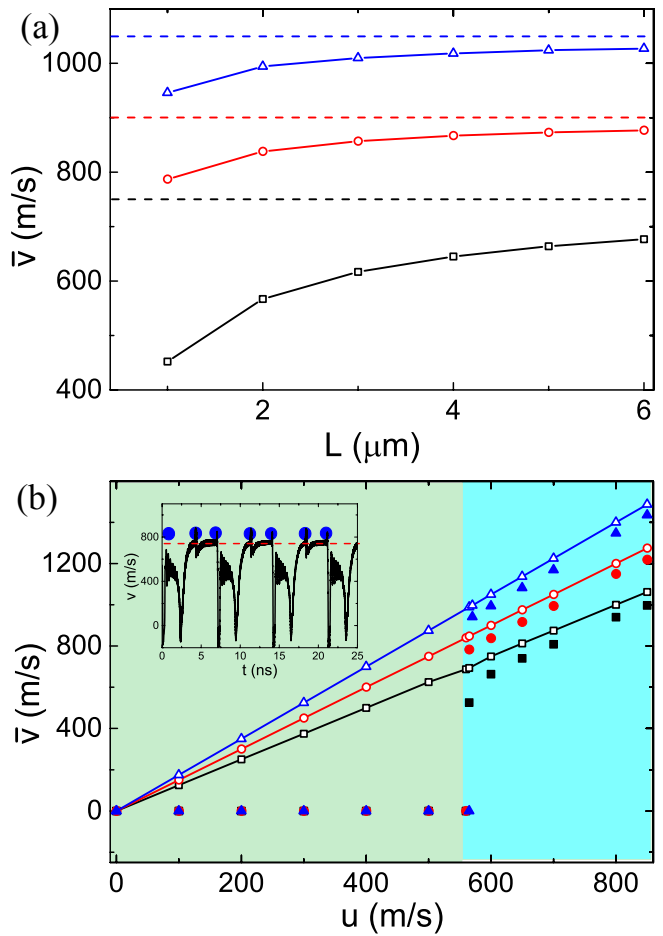}\\
\caption{(color online)
(a) $L-$dependence of $\bar v$ for $u=600$ m/s and $\beta=0.025$ (squares),
0.03 (circles), and 0.035 (up-triangles), all larger than $\alpha =0.02$.
The dash lines are $\beta u/\alpha$.
(b) $u-$dependence of $\bar v$ for $L=2\ \mu$m.
Fill symbols (squares for $\beta = 0.025$, circles for $\beta = 0.03$, and
up-triangles for $\beta = 0.035$) are numerical data in notched wire of
$w=48$ nm and $d=16$ nm.
Open circles are DW velocity of the corresponding homogeneous wire.
Straight lines are $\beta u/\alpha$.
Inset: instantaneous DW velocity
for $u=600$ m/s and $\beta = 0.025$.
The blue dots denote the moments when the DW is at notches.}
\label{fig3}
\end{figure}

\noindent\underline{Hindering phase:}
Things are quite different for $\beta > \alpha$.
Figure \ref{fig3}a shows that $\bar v$ increases monotonically with $L$
for $\beta = 0.025$, 0.03 and 0.035, which are all larger than $\alpha$.
In order to make a fair comparison with the results of $\beta<\alpha$,
Fig. \ref{fig3}b is the current density dependence of $\bar v$ for $L=2$
$\mu$m and $\beta = 0.025$ (filled squares), 0.03 (filled circles) and
0.035 (filled up-triangles). Again, DW velocities in the corresponding
homogeneous wires are presented as open symbols.
Take $\beta = 0.025$ as an example, although the average velocity jumps
at the depinning current density 565 m/s, it's still well below the DW
velocity in the corresponding uniform wire. The inset of Fig. \ref{fig3}b
shows the instantaneous DW velocity for $u=600$ m/s.
An antivortex is generated at the first notch. In contrast to the case of
$\beta<\alpha$, the antivortex slows down DW propagation velocity below
the value in the corresponding uniform wire. Moreover, the transverse wall
keeps its original chirality unchanged when the antivortex is annihilated
at wire edge, and no vortex/antivortex is generated at the second notch.
However, another antivortex is generated at the third notch.
This is the typical cycle of phase C1. 
As $u$ increases above 640 m/s, phase C1 disappears
and the DW passes all the notches without generating any vortices.
This motion is
termed as phase C2. For $\beta > 0.025$, only phase C2 is observed.
In C2, DW profile is not altered, and the average DW velocity is
slightly below that in a uniform wire.

\subsection{Transverse walls in very narrow wires}
\begin{figure}
\centering
\includegraphics[width=0.48\textwidth]{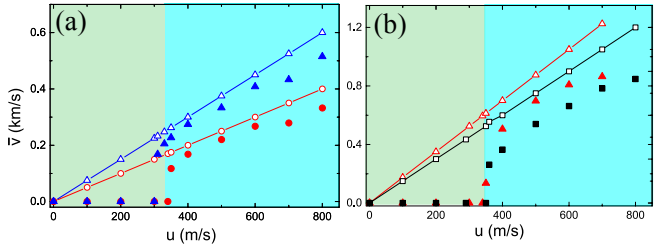}\\
\caption{(color online) (a) $u-$dependence of $ \bar v$ for
$\beta = 0.01$ (filled circles) and 0.015 (filled up-triangles).
(b) $u-$dependence of $ \bar v$ for $\beta = 0.03$ (filled squares) and 0.035
(filled up-triangles). Open symbols are DW velocity in the corresponding
homogeneous wires. Straight lines are $\beta u /\alpha$.
$ \bar v$ is below $\beta u/\alpha$ when $u> u_d$.
The nanowire is 8 nm wide and 1 nm thick while
the notch size is 10 nm wide and 2 nm deep for (a) and 50 nm wide
and 2 nm deep for (b). The separation of adjacent notches is 100 nm.}
\label{fig4}
\end{figure}

One interesting question is whether notches can boost DW propagation
in very narrow wires such that the nucleation of a vortex/antivortex is
highly unfavorable. To address this issue,
Fig. \ref{fig4}a are $u-$dependence of the average DW velocity on a 8 nm
wide wire for $\beta<\alpha$ (circles for $\beta=0.01$ and up-triangles for
$\beta=0.015$) with (filled symbols) and without (open symbols) notches.
When notches are placed, notch depth is 2 nm, $L = 100$ nm, $w=10$ nm.
DW velocity in the corresponding homogeneous wire (open symbols)
follows perfectly with $\bar v = \beta u/\alpha$ (straight lines).
It is clear that averaged DW velocity in the notched wire (filled symbols)
is below the values of the DW velocity in the corresponding homogeneous wire.
Take $\beta=0.015$ as an example, $\bar v$ is zero below $u_d = 310$ m/s
and jumps to an average velocity $\bar v \simeq 168$ m/s at $u_d$, which
is below the DW velocity in the corresponding uniform wire.

Things are similar for $\beta > \alpha$. Figure \ref{fig4}b is the
current density dependence of  $\bar v$  for $\beta = 0.03$
(filled squares) and 0.035 (filled up-triangles). Again, DW velocities in
homogeneous wire are presented as open symbols for a comparison.
The averaged DW velocity in the notched wire (filled symbols) is below
the values of the DW velocity in the corresponding homogeneous wire.

\begin{figure}
\centering
\includegraphics[width=0.48\textwidth]{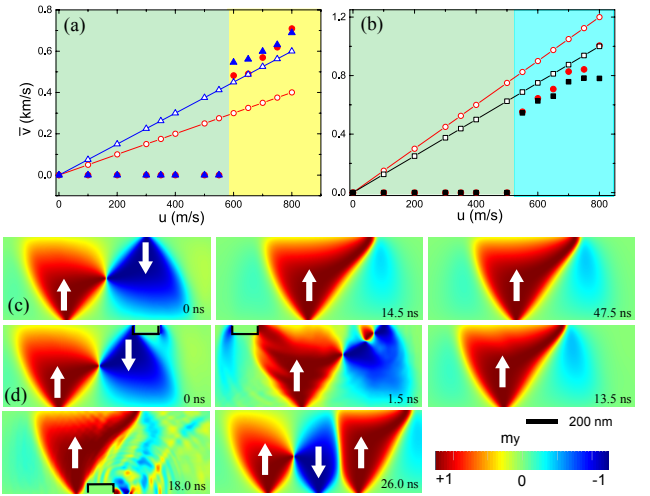}\\
\caption{(color online)
(a) $u-$dependence of $ \bar v$ for $\beta = 0.01$ (filled circles)
and 0.015 (filled up-triangles). (b) $u-$dependence of $ \bar v$
for $\beta = 0.025$ (filled squares) and 0.03 (filled circles).
Open symbols are DW velocity in the corresponding homogeneous wires.
Straight lines are $\beta u /\alpha$. $ \bar v$ is above (below)
$\beta u/\alpha$ when $u> u_d$ and $\beta < \alpha$ ($\beta > \alpha$).
The nanowire is 520 nm wide and 10 nm thick while the rectangular notch
is 160 nm wide and 60 nm deep. The separation of adjacent notches is 8 $\mu$m.
(c) and (d) The spin configurations in a uniform wire (a) and
in a notched wire (b) at various moments for $\beta = 0.01$ and $u=650$ m/s.
The time is indicated on the bottom-right corner of each configuration.
The color codes the value of $m_y$ and color bar is shown in the bottom-right
corner.}
\label{fig5}
\end{figure}

\subsection{Vortex walls in very wide wires}
Although our main focus is on transverse walls, it should be interesting
to ask whether DW propagation boost can occur for vortex walls.
It is well-known that a vortex/antivortex wall is more stable for a much
wider wire in the absence of a field and a current \cite{McMichael}.
One may expect that DW propagation boost would not occur in such a wire
because the boost comes from vortex/antivortex generation near notches
and a such vortex/antivortex exists already in a wider wire even in the
absence of a current. However, DW propagation boost was still observed
as shown in Fig. \ref{fig5} for a wire of 520 nm wide and 10 nm thick.
Rectangular notches of 60 nm deep and 160 nm wide are separated by $L=8\ \mu$m.
When $\beta<\alpha$ (Fig. \ref{fig5}a: circles for $\beta=0.01$ and
up-triangles for $\beta=0.015$), the average DW propagation velocities in
the notched wire (filled symbols) is higher than the DW velocity in the
corresponding homogeneous wire (open symbols) when $u>u_d$.
Figure \ref{fig5}b shows that the average DW propagation velocities in
a notched wire (filled symbols) is lower than that in the corresponding
homogeneous wire (open symbols) for $\beta>\alpha$ (squares for 
$\beta=0.025$  and circles for $\beta=0.03$). Figure \ref{fig5}c shows 
the spin configurations of the DW in the homogeneous wire of $\beta
= 0.01$ before a current is applied (the left configuration) and during
the current-driven propagation (middle and right configurations). 
When a current $u=650$ m/s is applied at 0 ns, a vortex wall moves downward.
The vortex was annihilated at wire edge, and the vortex wall transform
into a transverse wall. The DW keeps its transverse wall profile and
propagates with velocity of $\beta u/\alpha$ (solid lines in Fig.
\ref{fig5}a and \ref{fig5}b). The middle and right configurations are two
snapshots at 14.5 ns and 47.5 ns. Time is indicated in the bottom-right
corner. Figure \ref{fig5}d are snapshots of DW spin configurations in the
notched wire of $\beta=0.01$ when a current $u=650$ m/s is applied at $t=0$ ns.
At $t=0$ ns, a vortex wall is pinned near the first notch. Right after the
current is turned on, the vortex wall starts to depin and complicated
structures may appear during the depinning process as shown by the snapshot
at $t=1.5$ ns. At $t=13.5$ ns, the DW transforms to a transverse wall
and propagates forward. When the transverse wall reaches the second notch at
about $t= 18.0$ ns, new vortex core nucleates near the notch and drags
the whole DW to propagate forward. In contrast to the case of homogeneous
wire where a propagating DW prefers a transverse wall profile, DW with more
than one vortices can appear as shown by the snapshot at $t = 26.0$ ns.
The vortex core in this structure boosts DW velocity above the average DW
velocity of a uniform wire. This finding may also explain a surprising
observation in an early experiment \cite{Hayashi2006} that depinning
current does not depend on DW types.
A vortex wall under a current transforms into a transverse wall before
depinning from a notch. Thus both vortex wall and transverse wall have
the same depinning current.

\section{Discussion}
\subsection{Depinning process analysis}
\begin{figure}
\centering
\includegraphics[width=0.48\textwidth]{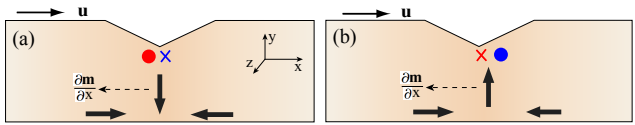}\\
\caption{(color online) Directions of vortex core magnetization (red symbols)
and non-adiabatic torque (blue symbols) for a clockwise transverse wall (a)
and a counterclockwise transverse wall (b). The dots (crosses) represent
$\pm z$-direction.}
\label{fig6}
\end{figure}

Empirically, we found that vortex/antivortex polarity is uniquely
determined by the types of transverse wall and current direction.
This result is based on more than twenty simulations that we have
done by varying various parameters like notch geometry, wire width,
magnetic anisotropy, damping etc. Within the picture that DW
depinning starts from vortex/antivortex  nucleation, the
$\beta-$dependence of depinning current density $u_d$ can be understood
as follows. For a clockwise (counterclockwise) transverse wall and
current in $-x$ direction, $p=+1$ ($p=-1$), as shown in Fig. \ref{fig6}.
If one assumes that vortex/antivortex formation starts from the
vortex/antivortex core, it means that the core spin rotates into
$+z$-direction for a vortex of $p=1$. For a clockwise wall, $\beta$-torque
($\beta \mathbf{m} \times \frac{\partial \mathbf{m}}{\partial x}$) tends
to rotate core spin in $-z$-direction, as shown in Fig. \ref{fig6}a, so the
presence of a small $\beta$-torque tries to prevent the nucleation of vortices.
Thus, the larger $\beta$ is, the higher $u_d$ will be.
This may be the reason why the depinning current density $u_d$ increases
as $\beta$ increases.

\begin{figure}
\centering
\includegraphics[width=0.48\textwidth]{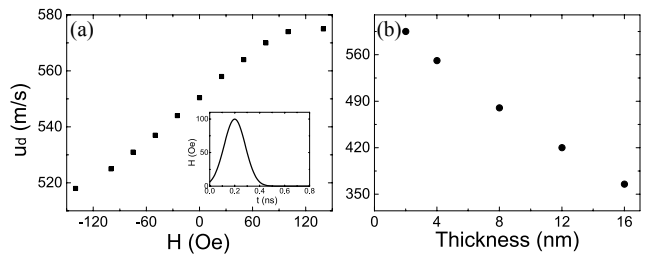}\\
\caption{(a) Depinning current density as a function of an external field.
A 0.4 ns field pulse in the $x$-direction is turned on simultaneously
with the current. The shape of a pulse of $H=100$ Oe is shown in the inset.
Since the depinning field of the wire (64 nm wide and 4 nm thick) is 150 Oe,
the field amplitude is limited to slightly below 150 Oe in the curve.
(b) Depinning current density as a function of nanowire thickness.}
\label{fig7}
\end{figure}
\vspace{6pt}

Our simulations suggest that DW depinning starts from vortex/antivortex
nucleation. Adiabatic spin transfer torque tends to rotate the spins at the
edge defect near a notch out of plane and to form a vortex/antivortex core.
Thus, any mechanisms that help (hinder) the creation of a vortex/antivortex
core shall decrease (increase) the depinning current density $u_d$.
To test this hypothesis, we use a magnetic field pulse of 0.4 ns along
$\pm x-$direction (shown in the inset of Fig. \ref{fig7}a) such that the
field torque rotates spins out of plane. Figure \ref{fig7}a is the numerical
results of the magnetic field dependence of the depinning current density
for a 64 nm wide wire with triangular notches of 48 nm wide and 16 nm deep.
The non-adiabatic coefficient is $\beta = 0.01$.
As expected, $u_d$ decreases
(increases) with field when it is along -$x-$direction (+$x-$direction)
so that spins rotate into +$z-$direction (-$z-$direction).
All other parameters are the same as those for Fig. 2.

If the picture is correct, one should also expect the depinning current
density depends on the wire thickness. The shape anisotropy impedes
vortex core formation because it does not favor a spin aligning in the
$z-$direction. The shape anisotropy decreases as the thickness increases.
Thus, one should expect the depinning current density decreases with
the increase of wire thickness. Indeed, numerical results shown in
Fig. \ref{fig7}b verifies the conjecture.
All other parameters are the same as that in Fig. \ref{fig7}a ($H=0$).
\begin{figure}
\centering
\includegraphics[width=0.48\textwidth]{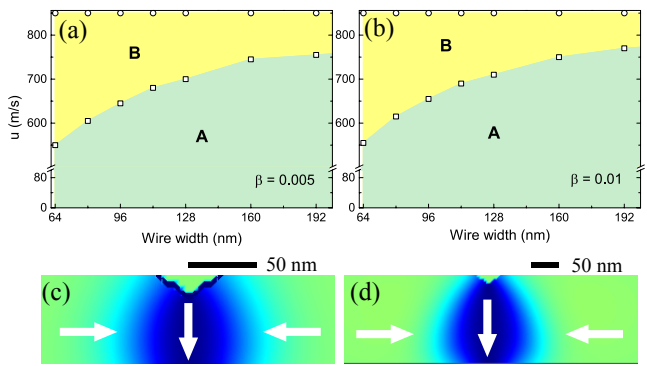}\\
\caption{(color online) (a) and (b) are nanowire width dependence of
depinning current density for $\beta = 0.005$
(a) and $\beta =0.01$ (b), respectively. The wire thickness is 4 nm and
notch size is fixed at $48 \times 16$ $\mathrm{nm^2}$.
(c) and (d) are the real configurations of initial domain walls
pinned at the notch for 64 nm and 160 nm wide wires, respectively.
The color coding is the same as that of Fig. \ref{fig5}.
The blue jagged lines indicate the profiles of triangular notches.}
\label{fig8}
\end{figure}

\subsection{Width effects on the depinning current density}

The DW propagating boost shown above is from the wire in which the notch
depth (16 nm) is relatively big in comparison with wire width (64 nm).
Naturally, one may ask whether the DW propagation boost exists also in
a wire when the notch depth is much smaller than the wire width.
To address the issue, we fix the notch geometry and vary the wire width.
Figure \ref{fig8} is the nanowire width dependence of depinning current
density when the notch size is fixed at $48 \times 16$ $\mathrm{nm^2}$.
Figures \ref{fig8}a and \ref{fig8}b show the phase boundary between
vortex-assisted boosting phase and the pinning phase. DW propagation
boost exists when nanowire width is one order of magnitude larger
than the notch depth. The top view of the wire and spin configurations
for 64 nm wide and 160 nm wide wires are shown in Fig. \ref{fig8}c
and Fig. \ref{fig8}d, respectively.

\subsection{DW Propagation and vortex dynamics}
DW propagation boost and slow-down by vortices can be understood from
the Thiele equation \cite{Thiele1973, Huber1982, Thiaville2005},
\begin{equation}
\mathbf{F}+\mathbf{G}\times(\mathbf{v-u})+\mathbf{D}\cdot(\alpha\mathbf{v}
-\beta \mathbf{u})=0,
\label{thiele}
\end{equation}
where $\mathbf{F}$ is the external force related to magnetic field
that is zero in our case, $\mathbf{G}$ is gyrovector that is zero for a
transverse wall and $\mathbf{G}=-2\pi qplM_s/\gamma \mathbf{\hat{z}}$
for a 2D vortex wall, where $q$ is the winding number (+1 for a vortex
and -1 for an antivortex), $p$ is vortex polarity ($\pm 1$ for core
spin in $\pm z$ direction) and $l$ is the thickness of the nanowire.
$\mathbf{D}$ is dissipation dyadic, whose none zero elements for a
vortex/antivortex wall are $D_{xx}=D_{yy}=-2M_s Wl/(\gamma \Delta)$
\cite{Huber1982}, where $W$ is nanowire width and $\Delta$ is the
Thiele DW width \cite{Thiele1973}. $\mathbf{v}$ is the DW velocity.

\begin{figure}
\centering
\includegraphics[width=0.48\textwidth]{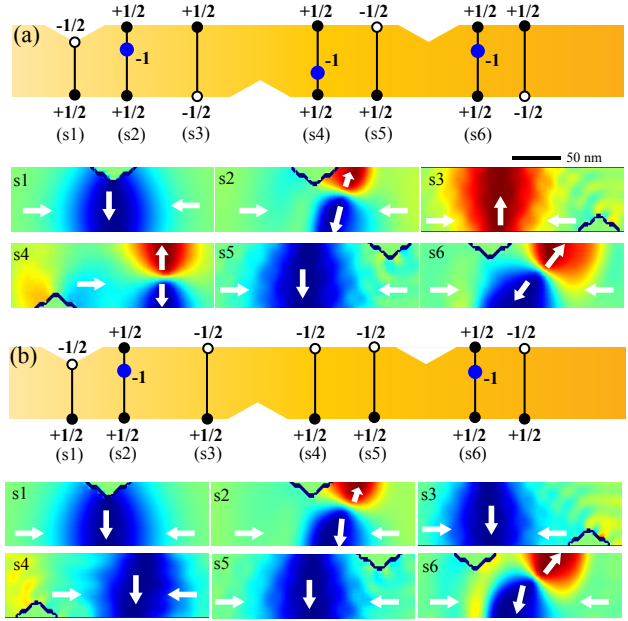}\\
\caption{(color online)
(a) Illustrations of changes of topological
defects (transverse DW edge defects and vortices) during the birth and
death of vortices in Phase B as a DW propagates from the left to the
right and the corresponding spin configurations at various moments.
Lines represent DWs. Big blue dots for
vortices and open circles for edge defects of winding number $-1/2$
and filled black circles for edge defects of winding number $1/2$.
The color coding is the same as that of Fig. \ref{fig5}.
The blue jagged lines indicate the profiles of triangular notches.
The nanowire is 64 nm wide and 4 nm thick.
The notch dimensions are $48 \times 16$ $\mathrm{nm^3}$. The interval
between adjacent notches is $L=1500$ nm. $u = 600$ m/s, $\beta=0.005$.
(b) Illustrations of changes of topological defects in Phase C1 and the
the corresponding spin configurations at various moments.
The nanowire is 64 nm wide and 4 nm thick.
The notch dimensions are $48 \times 16$ $\mathrm{nm^2}$. The interval
between adjacent notches is $L=2000$ nm. $u = 600$ m/s, $\beta=0.025$.}
\label{fig9}
\end{figure}

For a transverse wall, $\mathbf{v}=\beta\mathbf{u}/\alpha$ (solid lines)
agrees perfectly with numerical results (open symbols) in homogeneous
wires as shown in Figs. 2b and 3b without any fitting parameters.
For a vortex wall, the DW velocity is
%the Thiele equation becomes
%\begin{equation}
%\begin{aligned}
%&\alpha v_x - \frac{\pi q p \Delta}{W} v_y = \beta u \\
%&v_x +  \frac{\alpha W}{\pi q p \Delta} v_y = u. \\
%\end{aligned}
%\label{v}
%\end{equation}
%\sout{Thus DW velocity is}
\begin{equation}
\begin{aligned}
& v_y = \frac{1}{1+\alpha^2W^2/(\pi^2 \Delta^2)} \frac{W}
{\pi q p \Delta} (\alpha - \beta) u, \\
\end{aligned}
\label{vy}
\end{equation}
\begin{equation}
\begin{aligned}
v_x = \frac{u}{1+\alpha^2W^2/(\pi^2 \Delta^2)}\left(1
-\frac{\beta}{\alpha}\right ) + \frac {\beta u}{\alpha}.
\end{aligned}
\label{vx}
\end{equation}
$v_y$ depends on DW width, $\alpha$ as well as $\beta/\alpha$.
For a given vortex wall, $v_y$ has opposite sign for $\beta < \alpha$
and $\beta > \alpha$. In terms of topological classification of defects
\cite{Chern}, the edge defect of the transverse DW at the first notch
(Fig. 1a) has winding number $q=-1/2$, and this edge defect can only
give birth to an antivortex of $q=-1$ and $p=1$ while itself changes to
an edge defect of $q=1/2$ as shown in Fig. \ref{fig9}a. Empirically,
we found that antivortex polarity is uniquely determined by the types
of transverse wall and current direction. A movie visualizing the
DW propagation in boosting phase is shown in the Supplemental Movie 
\cite{Movie}.
All the parameters are the same as the inset of Fig. 2b.
The three segments of identical length 1200 nm are connected
in series to form a long wire.
%A detail analysis of this observation is presented in the Supplement.
%For a clockwise (counterclockwise)
%transverse wall and current in -x-direction, $p=1$ ($p=-1$).
%If one assumes that vortex formation starts from its core, it means
%that the core spin rotates into the +x-direction for a vortex of $p=1$.
%For a clockwise transverse wall, the $\beta$ term tends to rotate the
%core spin to the -z-direction. This may be the reason why the depinning
%current $u_d$ increases as $\beta$ increases.
When $\beta < \alpha$, the
antivortex moves downward ($v_y<0$) to the lower edge defect of winding
number of $q=1/2$. The lower edge defect changes its winding number to
$q=-1/2$ and the transverse DW reverses its chirality \cite{Yuan_jmmm2014}
when the vortex merges with the edge defect.
Then another antivortex of winding number $q=-1$ and $p=-1$ is generated at
the second notch on the lower wire edge and it moves upward ($v_y >0$).
The DW reverses its chirality again at upper wire edge when the antivortex dies.
Then this cycle repeats itself. The spin configurations corresponding to
various stages are shown in the lower panels of Fig. \ref{fig9}a.
When $\beta > \alpha$, as shown in Fig. \ref{fig9}b, the antivortex 
of $q = -1$ and $p = +1$ moves upward since $v_y > 0$. 
The chirality of the original
transverse wall shall not change when the antivortex is annihilated
at the upper edge defect because of winding number conservation.
No antivortex is generated at the even number notches and same type of the
antivortex is generated at odd number notches, hence the transverse wall
preserves its chirality throughout propagation. The corresponding
spin configurations are shown in the lower panels of Fig. \ref{fig9}b.
%\begin{figure}
%\centering
%\includegraphics[width=0.45\textwidth]{fig4.eps}\\
%\caption{(color online) Illustrations of changes of topological
%defects (transverse DW edge defects and vortices) during the birth and
%death of vortices in phase B (a) and phase C1 (b) as a DW propagates from
%the left to the right. Lines represent DWs. Big blue dots for
%vortices and open circles for edge defects of winding number $-1/2$
%and filled black circles for edge defects of winding number $1/2$.
%}
%\label{fig4}
%\end{figure}
%begin{figure}[!htb]

The second term in Eq. (3) (for $v_x$) is $\beta u/\alpha$, the same
as the transverse DW velocity in a homogeneous wire (straight
lines in Figs. 2b and 3b). The first term depends on DW properties as
well as $\beta$ and $\alpha$. It changes sign at $\beta=\alpha$.
$v_x$ is larger than $\beta u/\alpha$ in the presence of vortices if
$\beta < \alpha$. Therefore, in this case vortex generations and vortex
dynamics boost DW propagation.
For small $\alpha$ and to the leading order correction in $\alpha$ and
$\beta$, Eq. (3) becomes $v_x=u-(\alpha^2-\alpha\beta)uW^2/(\pi^2
\Delta^2)$. Thus, the longitudinal velocity equals approximately $u$
and depends very weakly on $\beta$. This is what was observed in Fig. 2b.
$v_x=u$ corresponds to the complete conversion of itinerant electron
spins into local magnetic moments. Although the Thiele equation
cannot explain why a DW generates vortices around notches in phase B,
it explains well DW propagation boost for $\beta<\alpha$.
This result is in contrast to the field-driven DW propagation
where vortex/antivortex generation reduces the Walker breakdown field
and inevitably slows down DW motion \cite{Milta,Yuan_jmmm2014}.

Before conclusion, we would also like to point out that it is possible to
realize both $\beta<\alpha$ (boosting phase) and $\beta>\alpha$ (hindering
phase) experimentally in magnetic materials like permalloy with damping 
coefficient engineering. 
A recent study \cite{Reidy2003} demonstrated that $\alpha$ of permalloy
can increase by four times through a dilute impurity doping of
lanthanides (Sm, Dy, and Ho).
\section{Conclusions}
In conclusion,
%notch effects on the current-induced DW propagation
%is investigated at relative low current density under which the DW
%motion is simple enough to be described. When $\beta<\alpha$, series
%of vortices with well defined topological number nucleate when a DW
%reaches notches. It is found that
notches can boost DW propagation when $\beta<\alpha$.
The boost is facilitated by antivortex generation and motion, and
boosting effect is optimal when two neighboring notches is separated
by the distance that an antivortex travels in its lifetime. In the boosting
phase, DW can propagate at velocity $u$ that corresponds to a complete
conversion of itinerant electron spins into local magnetic moments.
When $\beta > \alpha$, the notches always hinder DW propagation.
According to Thiele's theory, the generation of vortices increases
DW velocity for $\beta < \alpha$ and decreases DW velocity when
$\beta > \alpha$. This explains the origin of boosting phase and
hindering phase.
Furthermore, it is found that a vortex wall favored in a very wide
wire tends to transform to a transverse wall under a current. This may
explain experimental observation that the depinning current density
is not sensitive to DW types.
\section{acknowledgments}
We thank Gerrit Bauer for useful comments.
HYY acknowledges the support of Hong Kong PhD Fellowship.
This work was supported by NSFC of China (11374249) as well as
Hong Kong RGC Grants (163011151 and 605413).


\begin{thebibliography}{}
\bibitem{note1}OOMMF package was used in the early stage of this research.
In order to simulate a long and wide wire, we switched to MUMAX package.
Two packages give almost identical results on shorter wires,
and the results presented here were generated from MUMAX.
\bibitem{note2}Notch geometry affects depinning current because of the change
of current density and perpendicular shape anisotropy (see Ref. 22) in notch area.
Both effects help to generate vortices and thus reduce the depinning current.
This may explain the result.
\end{thebibliography}
\end{document}